\documentclass[preprint,12pt]{elsarticle}

\usepackage{amssymb}
\usepackage{ulem}
\usepackage{natbib}
\usepackage{amsmath}
\usepackage{color}

\usepackage{url}
\journal{}
\begin{document}

\begin{frontmatter}

\title{Using Networks and Partial Differential Equations to Predict Bitcoin Price}
\author[label1]{Yufang Wang}
\author[label2]{Haiyan Wang\corref{mycorrespondingauthor}}
\cortext[mycorrespondingauthor]{Corresponding author}
\ead{haiyan.wang@asu.edu}

\address[label1]{School of Statistics,
        Tianjin University of Finance and Economics, Tianjin 300222, China;}
 \address[label2]{School of Mathematical and Natural Sciences, Arizona State University, AZ 85069, USA;}

\begin{abstract}
Over the past decade, the blockchain technology and its Bitcoin cryptocurrency have received considerable attention. Bitcoin has experienced significant price swings in daily and long-term valuations. In this paper, we propose a partial differential equation (PDE) model on the bitcoin transaction network for predicting bitcoin price. Through analysis of bitcoin subgraphs or chainlets, the PDE model captures the influence of transaction patterns on bitcoin price over time and combines the effect of all chainlet clusters.  In addition, Google Trends Index is incorporated  to the PDE model to reflect the effect of bitcoin market sentiment.  The experiment shows that the average accuracy of daily bitcoin price prediction is 0.82 for 362 consecutive days in 2017. The results demonstrate the PDE model is capable of predicting bitcoin price. The paper is the first attempt to apply a PDE model to the bitcoin transaction network for predicting bitcoin price.
\end{abstract}


\end{frontmatter}

\section{Introduction}

Bitcoin is currently the world's leading cryptocurrency and the blockchain is the technology that underpins it. The concept of bitcoin was first suggested in 2008 by Satoshi Nakamoto, and it became fully operational in January 2009 \cite{nakamoto2008bitcoin}. By May 2018, the market
capitalization of Bitcoin had arrived at nearly \$115 billion dollars.   In contrast to the traditional financial asset, whose records of everyday monetary transactions are considered
highly sensitive and are kept private,
Bitcoin has no financial intermediaries and
a complete list of its transactions is publicly available
in a public ledger.
This publicly distributed ledger creates opportunities for people to
observe all the financial interactions on the blockchain network, and analyze how
the assets circulate in time.

Bitcoin value (i.e., the price of a bitcoin) often undergoes large swings over short periods.
For instance, at the beginning of 2013, the price
started at nearly \$13 per bitcoin 
and then rocketed to \$230 on 9 April, yielding almost 1700\% profits in less than four
months;
In 2017, the
value of a single bitcoin increased 2000\%,  going from \$863 on January 9, 2017, 
to a high of \$17,550 on December 11, 2017.

Given the high volatility and the difference from traditional currencies, the price of bitcoin is extremely difficult to predict.
A few studies have been conducted on prediction or estimation for bitcoin prices. Regression models are the method mostly used for bitcoin price prediction by considering some potential price-affecting factors. For instance, Ciaian et al. \cite{ciaian2016economics} predict bitcoin price with a linear regression model by considering some factors in marco-finance and attractiveness for investors. Jang et al.  \cite{jang2017empirical} introduce blockchain information (such as hash rate and block generation rate) to increase the prediction accuracy by
a Bayesian neural network. Researchers also apply machine learning techniques to make bitcoin price predictions \cite{velankar2018bitcoin,kurbucz2019predicting,chen2020bitcoin,atsalakis2019bitcoin}. Atsalakis et al. \cite{atsalakis2019bitcoin} use ``a hybrid neuro-fuzzy controller, namely PATSOS, to forecast the direction in the change of the daily price of bitcoin."
Cretarola et al. \cite{cretarola2018modeling} propose an ordinary difference equation to describe the behavior of bitcoin price by considering investors' attention for bitcoin. As the publicly available ledger of the Bitcoin network can be represented by a directed graph, some work exists to make bitcoin prediction through network analysis \cite{kurbucz2019predicting,akcora2018forecasting}.
Kurbucz et al. \cite{kurbucz2019predicting} predict the price of bitcoin by the most frequent edges of its
transaction network. This study shows the utility of global graph features can be used to predict
bitcoin price.


It is generally accepted that bitcoin price is significantly affected by attention to or sentiment about the Bitcoin system itself \cite{kristoufek2013bitcoin,bukovina2016sentiment,kristoufek2015main}. The frequency of searches for the term ``bitcoin" in Google Trends  has proved to be a good measure of interest \cite{cretarola2018modeling,kristoufek2013bitcoin,engelberg2011search}.
Cretarola et al. \cite{cretarola2018modeling} regard the Google Trends index as a proxy for the attention measure and thus propose a bivariate model in continuous time to
describe the behavior of bitcoin price.
Kristoufek et al.  \cite{kristoufek2013bitcoin} demonstrate quantitatively that
not only are the Google Trends Index
and the prices connected but a pronounced asymmetry also exists between the effect of increased
interest in the currency while it is above or below its trend value.

Recently, Akcora et al. \cite{akcora2018forecasting} studied the influence of local topological structures on bitcoin price dynamics.  They combined ``chainlets"  of the Bitcoin transaction networks with statistical models to predict bitcoin price.  Essentially, chainlets are special forms of network motifs or subgraphs in the address-transaction bitcoin graph.  Chainlets describe transactions occurring in a blockchain and each chainlet represents a trading decision or transaction pattern.  In \cite{akcora2018forecasting} Akcora et al. find that certain types of chainlets exhibit an important role in bitcoin price prediction. It has been found that bitcoin price is mainly and strongly linked with transaction activities \cite{kristoufek2015main,koutmos2018bitcoin}. Analysis of motifs of the bitcoin transaction networks has been found to be an indispensable tool to unveil hidden mechanisms of Bitcoin networks for the bitcoin price dynamics.


In this paper, we aim to propose a partial differential equation (PDE) model to predict bitcoin price.  As indicated in \cite{akcora2018forecasting}, the predictive utility of different types of transactions for the bitcoin price dynamics may be different. As a result, we apply spectral analysis to aggregate bitcoin transaction subgraphs (chainlets) into clusters with similar types of transaction patterns.  We embed the clusters of chainlets into a Euclidean space and develop a PDE model to incorporate the bitcoin market sentiment with the Google Trends Index.  The framework of PDE models developed by the authors for information diffusion in online social networks in \cite{Haiyan2019book,wang2018prediction, wang2016regional} is adapted to the Bitcoin transaction network for characterizing the effect of the chainlet clusters. The PDE model enables us to describe the influence of the clusters on the price over time. The experiment shows that the average accuracy of daily bitcoin price prediction is 0.82 for 362 consecutive days in 2017. The results demonstrate the PDE model is capable of predicting bitcoin price. The paper is the first attempt to apply a PDE model on bitcoin transaction network for predicting bitcoin price.

In summary, the new contributions of this paper are as follows: (i) it is the first attempt to establish a PDE model on the bitcoin transaction network for the bitcoin price dynamics; (ii) spectral analysis of the bitcoin transaction network is applied to obtain chainlet clusters to represent different transaction patterns and predictive utilities; (iii)  within the framework of the PDE prediction model, we integrate Google Trends and chainlets to account for the effect of chainlet clusters; (iv) the average prediction accuracy of this model is 0.82 for 362 consecutive days in 2017.

The remainder of the paper is organized as follows: some basic concepts are introduced in Section \ref{chianletsAndBitcoin}. The PDE based prediction model is proposed in Section \ref{models}. The prediction process is described in Section \ref{prediction}.
Section \ref{end} concludes the paper with discussions.

\section{Bitcoin and chainlet}
\label{chianletsAndBitcoin}
A blockchain is a distributed ledger that records transactions in blocks without requiring a trusted central authority. Each block contains a set of transactions and it has a hash link to its previous
block, thus creating a chain of chronologically ordered blocks. When transactions happen at the same time, they will be recorded in the same block.

Bitcoin addresses are used for receiving
bitcoins. A transaction represents the flow of bitcoins
from input address to output addresses over time. A transaction is multi-input, multi-output, which means
that a transaction may have more than one
input address  and
more than one output address.
Users take part in the bitcoin economy through addresses and a user can have two or more addresses at the same time.
\begin{figure}[!h]
\centering
\includegraphics[width=0.95\textwidth]{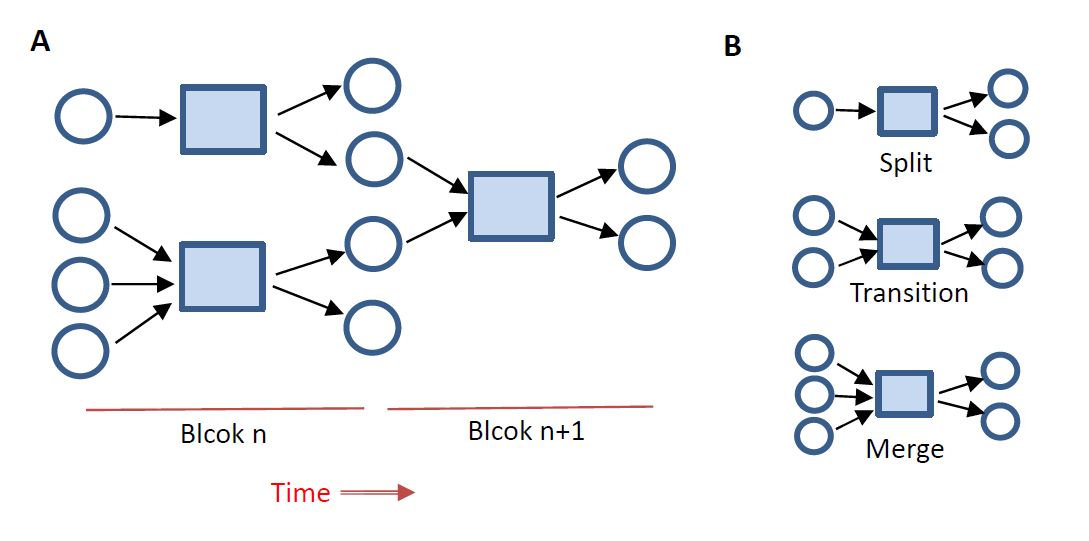}
\caption{\textbf{(A)} A transaction-address graph; \textbf{(B)} Split ($C_{1\longrightarrow 2}$),  Transition ($C_{2\longrightarrow 2}$) and Merge ($C_{3\longrightarrow 2}$) Chainlets. The three types, Merge, Transtion and Split are determined according to the relative number of input addresses and output addresses, and correspond to the state that  the former is greater than, equal to, or less than the latter respectively. Addresses and
transactions are shown with circles and rectangles,
respectively. An arrow indicates a transfer of
bitcoins.}
\label{bitcoin}
\end{figure}

The transaction-address graphs \cite{akcora2018forecasting,akcora2018bitcoin} is a directed graph, which is vital important for knowing bitcoins' flowing state. In this graph, the set $\{Address,Transaction\}$ consists of the vertexes and an edge represents the bitcoin's transfer between and address node and a transaction node as in figure  \ref{bitcoin}.
As transactions are the only means to manage bitcoins, so
bitcoins can be divided or aggregated only by being spent.
For instance, a transaction involves multiple addresses and every user can have different addresses, so the user can use a transaction to split, merge
or move bitcoins between its own addresses. Therefore, each transaction with its input and output addresses represents a decision, encoded by a subgraph in  the transaction-address graph.

The subgraph, composed of a transaction with its input and output addresses, is called \textit{1-chainlet}, or simply may be called \textit{chainlet}, which is a novel graph data model, proposed in 2018 \cite{akcora2018forecasting} for studying bitcoin price dynamics. A chainlet with $x$ inputs and $y$ outputs is often noted as $C_{x\longrightarrow y}$. Chainlets have distinct shapes that reflect their role in the network.  They are the building block in blockchain analysis; they represent different transaction patterns, and reflect different decisions.
In this paper, we analyze bitcoin chainlets and predict bitcoin price dynamics.

\section{PDE model based on chainlets and Google Trends}
\label{models}

In this section, we establish a PDE model for bitcoin price prediction. This PDE model is based on chainlets and the Google Trends Index for the bitcoin price. Different types of chainlet clusters represent different transaction patterns and transaction decisions;  thus, they provide different predictive utility for bitcoin price. The PDE model proposed below captures the influence of the chainlet clusters and combines the effects of all the clusters for predicting bitcoin price.



\subsection{Chainlets clustering}
In this paper we use chainlets as the building block for bitcoin price prediction which represents a type of immutable decision. To make a better prediction, we use spectral clustering to aggregate chainlets into clusters with similar type of transactions.

\textbf{Chainlet Network} A transaction with inputs and outputs composes a chainlet. Just as in \cite{akcora2018forecasting}, we denote $C_{x\longrightarrow y}$ as a chainlet if this chainlet has $x$ inputs and $y$ outputs. Different chainlets have different values of  $x$ and $y$ ($x$ and $y$ are positive integers). Though the Bitcoin protocol limits the number of input and output addresses for a transaction, the number of inputs and outputs can still reach thousands. As a result, millions of different chainlets occur (e.g. $C_{2000\longrightarrow 20}$ and $C_{2000\longrightarrow 120}$). However, an analysis of the entire bitcoin history shows that 97.57\% of the chainlets have fewer than 20 inouts and outputs \cite{akcora2018forecasting}.
This means that in bitcoin blocks, a sufficiently large number of chainlets satisfy $1\leq x<20, 1\leq y < 20$.

Therefore, we build a weighted graph $G(V,E)$ with 400 nodes, where each node in $V$ is a kind of chainlet or chainlet set and the graph node set
 $\{\mathbb{C}_{x\longrightarrow y}, 1\leq x \leq 20, 1\leq y \leq 20\}$ is defined as
\begin{equation*}
   \mathbb{C}_{x \longrightarrow y} =
   \begin{cases}
	   C_{x\longrightarrow y},& \text{if } x<20\quad \text{and}\quad y<20; \\
	 \{C_{x\longrightarrow j}, 20\leq j<+\infty \},& \text{if }x<20\quad\text{and}\quad y=20;\\
      \{C_{i \longrightarrow y}, 20 \leq i<+\infty \},& \text{if }x=20\quad\text{and}\quad y<20;\\
\{C_{i\longrightarrow i}, 20 \leq i<+\infty, 20\leq j<+\infty  \},& \text{if }x=20\quad\text{and}\quad y=20.
  \end{cases}
\end{equation*}
An edge $i$ and $j$ in $E$ and its weight are determined by the Pearson Correlation Coefficient of relative historical daily transaction volume, with a correlation cut threshold of $\theta$.

\textbf{Chainlets Clustering}
Different transactions (transaction with distinct inputs and outputs)  contribute differently for the bitcoin price formation. Four hundred distinct chainlets exist in the chainlet network $G$ and each chainlet represents a type of transaction form. However, intuitively, it is not necessary to distinguish all chainlets. For example,  the transaction patterns $\mathbb{C}_{3\longrightarrow 1}$ and $\mathbb{C}_{4\longrightarrow 1}$ have no meaningful difference. Therefore, we cluster chainlets by spectral clustering, which divides the graph by using the eigenvectors of Laplacian matrix \cite{von2007tutorial}.
Thus, each chainlet cluster represents certain similar types of transaction pattern, which may have different predictive utilities for the bitcoin price.  Our PDE model proposed in the next subsection is capable of capturing the influence of these chainlet clusters and integrating the effects of all clusters for bitcoin price prediction.

\begin{figure}[!h]
\centering
\includegraphics[width=0.95\textwidth]{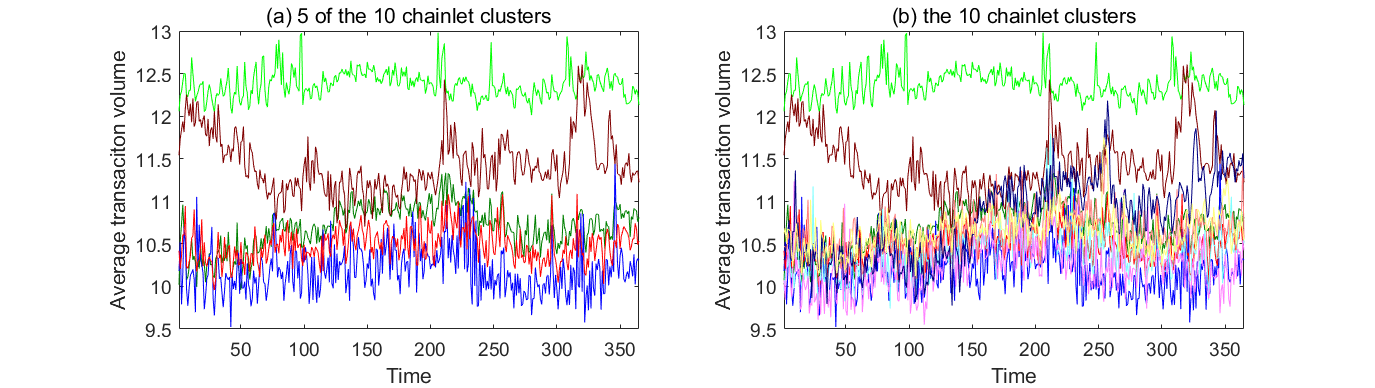}
\caption{(a)Average transaction volumes for 5 of the 10 chainlet clusters; (b) The 10 chainlet clusters obtained by the spectral clustering method.  Each color represents a cluster. The time period is from January 1, 2017 to December 31, 2017.}
\label{volumes}
\end{figure}

In this paper, we apply the daily transaction volumes of the 400 kinds of chainlets
from December 1, 2016 to December 30, 2016 (denoted here it as Data-set 1)
to build the chainlet network $G$ with a Pearson Correlation cut threshold $\theta=0.6$.
In fact, we can use data in other time periods
to build network as well, as long as the time interval of Data-set 1 is earlier
than the period for which we want to make prediction for bitcoin price. For example, if
we want to predict the bitcoin price in February,  2017, we can first build the
network based on the data in January, 2017.
We obtain 10 chainlet clusters by applying spectral clustering method to the above chainlet network $G$. Our prediction period is from January 1, 2017, to December 31, 2017 (denoted it as Data-set 2).  Figure  \ref{volumes} shows  the average transaction volumes of all the chainlet clusters for this period. 

Bitcoin price is mainly and strongly linked with transaction activities, especially transaction volumes \cite{kristoufek2015main,koutmos2018bitcoin}.
Figure  \ref{volumes} shows that the chainlet clusters we obtain have different levels of transaction volumes.
\subsection{Modeling bitcoin price with PDE}
In this section, we develop a PDE model to model the influence of chainlet clusters with different transaction patterns and combine their effects on the bitcoin price, for the purpose of predicting the bitcoin price.




We apply the authors' framework of PDE models for information diffusion in online social networks \cite{Haiyan2019book,wang2018prediction, wang2016regional} to the Bitcoin transaction network for characterizing the influence of the chainlet clusters on bitcoin price.

%

\begin{figure}[!h]
\centering
\includegraphics[width=0.8\textwidth]{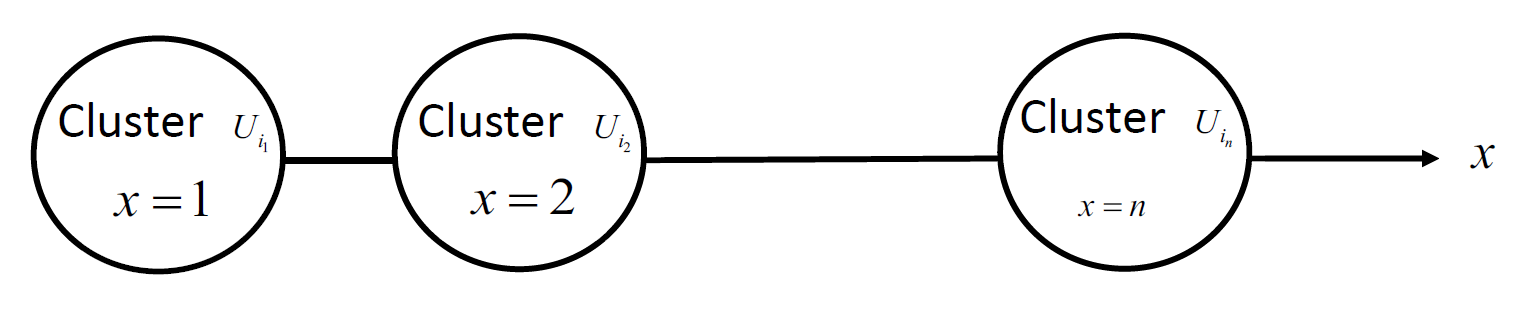}
\caption{Embedding of chainlet clusters into the $x$-axis.}
\label{clusters}
\end{figure}

To apply a PDE model to the interaction of the chainlet clusters, one must embed the chainlet clusters $U_1, U_2,\ldots, U_n$ ($n$ is the number of chainlet clusters) into a Euclidean space and arrange them in a meaningful order.  In this paper, the chainlet clusters are mapped  onto a line such that connected clusters stay as close as possible.  Specifically, we now treat each chainlet cluster
as a node in a new graph $G_{new}$ where the strength of edges is the summation of all
weights between two clusters.   The Fiedler vector (the eigenvector corresponding to the second smallest eigenvalue) of the Laplacian matrix of the new graph $G_{new}$ can map these chainlet clusters onto a line,  showing the order as $U_{i_1}, U_{i_2},\ldots, U_{i_n}$  \cite{Haiyan2019book}. On this line, connected nodes stay as close as possible, ensuring that the continuous model can capture the influence of the chainlet clusters.

Having embedded chainlet clusters to the Euclidean space,  let $u(x,t)$ represent the effect of chainlet cluster $x$ on the bitcoin price.  The formulation of the spatio-temporal model for the bitcoin network follows the balance law: the rate at which a given quantity
changes in a given domain must equal the rate at which it flows across its
boundary plus the rate at which it is created, or destroyed, within the domain.  The PDE model can be conceptually divided into two processes: an internal process within each chainlet cluster and external process among chainlet clusters. Similar derivation for the PDE model has been used in our previous work for PDE models for information diffusion in online social networks in \cite{Haiyan2019book,wang2018prediction,wang2016regional}.
Our proposed PDE-based model is
\begin{equation}
\label{new_model}
\begin{split}
\frac{\partial u(x,t)}{\partial t}=\frac{\partial }{\partial x}\big(d(x)\frac{\partial u(x,t)}{\partial x}\big)+r(t)u(x,t)h(x),
\end{split}
\end{equation}
where $r(t)u(x,t)h(x)$ describes the rate of change of bitcoin price within the cluster $x$; $r(t)$ represents the rate of change with respect to $t$; $h(x)$ describes the spatial heterogeneity of different chainlet clusters or transaction patterns. $\frac{\partial}{\partial x} \left(d(x)\frac{\partial u(x,t)}{\partial x}\right)$ reflects the rate of change of the bitcoin price among chainlet clusters, $d(x)$ describes the interaction of chainlet clusters.  Because $u(x,t)$ represent the influence of chainlet cluster $x$ on the bitcoin price, we have
\begin{equation}
\label{model00}
 \text{predicted bitcoin price}= \int u(x,t) dx
\end{equation}

%

Investor sentiment about bitcoin, which can be represented by the Google Trends Index,  is a key factor for determining bitcoin price. It is plausible to assume that
\begin{equation}
\label{change}
u(x,t)\equiv b_0 m(x,t)+\alpha(x),
\end{equation}
where $m(x,t)$ is the predictive utility of the Google Trends Index of chianlet cluster $x$;
$ b_0$ is a scale factor to bitcoin price.  $\alpha(x)$ describes the heterogeneity of different chainlet clusters on bitcoin price.  In this way,  chainlet clusters with larger trading volumes do not necessarily have more influence on the bitcoin price.

We derive a PDE model with respect to $m(x,t)$. Note both $h(x)$ and $\alpha(x)$ represent the spatial heterogeneity of different chainlet clusters and therefore we can assume that $h(x)=k\alpha(x)$ with a constant $k$.  Substituting (\ref{change}) for  (\ref{new_model}) and (\ref{model00}) and  assuming  $d(x)\equiv d>0$, $d$ is a constant, the prediction model of bitcoin price is

\begin{equation}
\label{model2}
   \begin{cases}
\frac{\partial m(x,t)}{\partial t}=d\frac{\partial^2 m }{\partial x^2}+k\alpha(x)r(t)\Big(m(x,t)+\frac{1}{ b_0}\alpha(x)\Big)+\frac{d}{b_0}\alpha^{''}(x),
\\
m(x,1)=\phi(x),L_1<x<L_2,
\\
\frac{\partial m}{\partial x}(L_1,t)=\frac{\partial m}{\partial x}(L_2,t)=0,t>1,\\
 \text{Predicted bitcoin price at time}\ t=\int_{L_1}^{L_2}\Big( b_0m(x,t)+\alpha(x)\Big) dx,
  \end{cases}
\end{equation}
where
\begin{itemize}
  \item $\alpha(x)$ satisfies $\alpha(x_i)=\alpha_i, i=1,2,\ldots,m$, where $\alpha_i$ is the heterogeneity of the cluster at location $x_i$ and $n$ is the number of clusters; and
      $r(t)$ satisfies exponential decay with time;
      In this work,
  $r(t)$ is assumed the form of  $r(t)=b_1+e^{-(t-b_2)}$;
  we construct $\alpha(x)$ by the cubic spline interpolation \cite{Gerald1994book} with the condition of $\frac{\partial \alpha}{\partial x}(L_1)=\frac{\partial \alpha}{\partial x}(L_2)=0$. Therefore, the second derivative $\alpha^{''}(x)$ exists and is continuous.

\item
      Neumann boundary condition $\frac{\partial m}{\partial x}(L_1,t)=\frac{\partial m}{\partial x}(L_2,t)=0, \quad t>1$ is applied and it has been assumed  no flux of information flow across the boundaries at $ x=L_1, L_2$; Initial function $m(x,1)=\phi(x)$ describes the influence of every chainlet cluster and it is constructed from the historical data by cubic spline interpolation.
  \item Parameters $d,b_0,b_1,b_2, \alpha_i, i=1,2,\ldots,n$ are determined by the known historical data of $m(x_i,t_j)$;
  \item The historical  predictive utility of the Google Trends Index  on bitcoin price
$$m(x_i,t_j), \ i=1,2,\ldots,n; j=1,2,\ldots,N$$
can be obtained by computing
\begin{equation*}
m(x_i,t_j)=\text{(Google Trends Index on ``Bitcoin" at time}\ t_i) * P_0,
\end{equation*}
where
\begin{equation*}
P_0=\frac{\text{bitcoin transaction volume of chianlet cluster $ x_j$ at time $t_i$}}{\text{total bitcoin transaction volume of all chianlet clusters  at $\ t_i$}}
\end{equation*}
%

\end{itemize}
\section{Prediction of bitcoin price}
\label{prediction}

\subsection{Data}
As mentioned in the introduction, bitcoin price is related to the transaction volumes and the investors' attention that can be measured by Google \cite{bukovina2016sentiment}.
The daily bitcoin transaction volumes of all transactions and the daily bitcoin price (here it refers to the intraday open price, denominated in USD in our work)  are downloaded from \url{https://github.com/cakcora/coinworks.} from January 1, 2017 to December 31, 2017.
These data sets are extracted from the original bitcoin data, which are all publicly available at the Bitcoin Core page (\url{https://bitcoin.org/en/download.}).
The Google Trends Index on ``bitcoin" captures the attention of
retail/uniformed investors \cite{engelberg2011search}. These data can be obtained from \url{https://trends.google.it/trends/?geo=IT} with a keyword ``bitcoin" .

\subsection{Prediction process and results}
To predict the bitcoin price at time $t_{N+1}$, the prediction process consists in determination of parameters in (\ref{model2}) using known historical data $\{m(x_i,t_j), i=1,2,\ldots, n; j=1,2,\ldots,N\}$, and solving the PDE-model (\ref{model2}) to  make a one-step prediction for $m(x,t_{N+1}), x\in[L_1,L_2]$. Thus, the predicted bitcoin price at time $t_{N+1}$ is given by
\begin{equation}
\label{price}
Price(t_{N+1})=\int_{L_1}^{L_n}\Big(hm(x,t_{N+1})+\alpha(x)\Big) dx,
\end{equation}

Specifically, we combine a tensor train (TT) global optimization approach
\cite{oseledets2011tensor} and Nelder{Mead simplex local optimization method \cite{lagarias1998convergence} to train the
PDE parameters. After each determination of the model parameters, we apply the fourth-order Runge-Kutta algorithm to compute the PDE
for one-step forward in time dimension numerically.

In this work, we obtain 10 chainlet clusters, therefore, $n=10, L_1=x_1=1, L_{10}=x_{10}=10$.  We use 3 day historical data of $m(x_i,t_j)$  to predict the the bitcoin price of the 4th day. We make a prediction for the whole year of 2017 and the prediction results cover 362 days from January 4, 2017.
The relative accuracy (RA), defined as $$\text{RA}=1-\frac{|P_{real}-P_{predict}|}{P_{real}}$$ is used to measure the prediction accuracy,
 where $P_{real}$ is the real bitcoin price  at every data collection
time point and $P_{predict}$ is the predicted bitcoin price.

\begin{figure}[!h]
\centering
\includegraphics[width=0.95\textwidth]{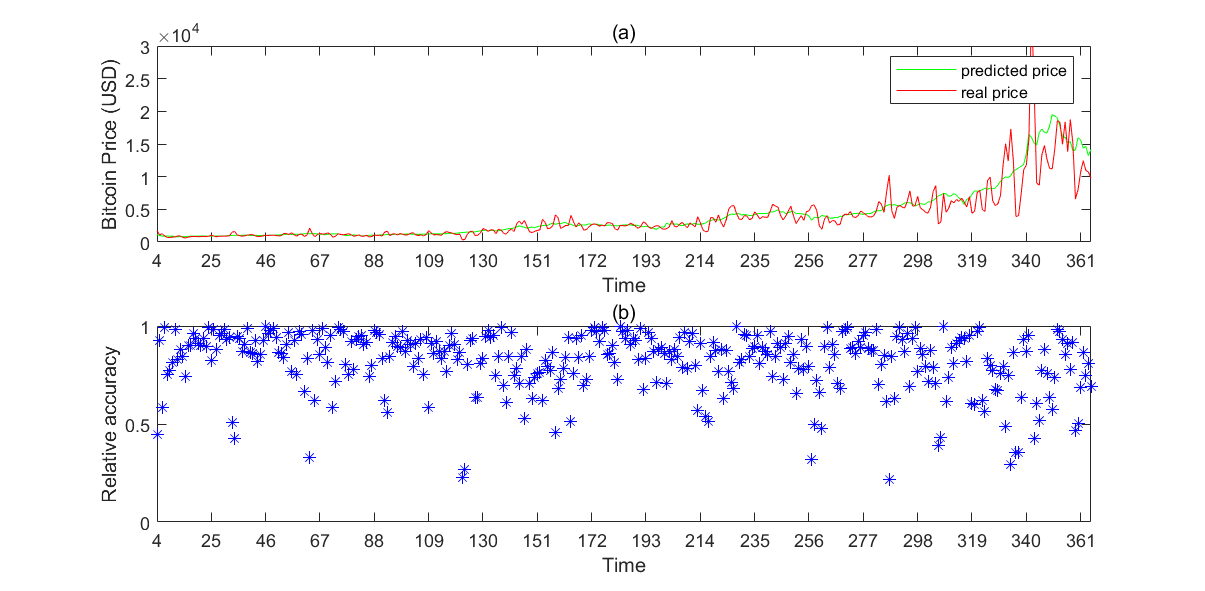}
\caption{The relative accuracy for bitcoin price prediction from January 4, 2017 to December 31, 2017. Here the relative accuracy (RA) is the conventional definition as $\text{RA}=1-\frac{|P_{real}-P_{predict}|}{P_{real}}$,  where $P_{real}$ is the real bitcoin price  at every data collection
time point and $P_{predict}$ is the predicted Bitcoin price through  (\ref{model2}).}
\label{RA}
\end{figure}
\begin{table}[!htbp]
  \centering
    \begin{tabular}{cc|cc}
    \hline
    Total days & 362   & Average Relative Accuracy (RA)=0.82 \\
    \hline
     days of RA$>$0.9 & 134   & (days of RA$>$0.9)/(total days) = 37\% \\
    days of RA$>$0.8  & 237   & (days of RA$>$0.8)/(total days) = 65\% \\
    days of  RA$>$0.7& 296   & (days of  RA$>$0.7)/(total days) = 82\% \\
    \hline
    \end{tabular}%
      \caption{The statistics of days for the prediction results of bitcoin price from January 4, 2017 to
December 31, 2017.  362  and 134 mean that during the prediction period of 362 days, there are 134 days whose relative accuracy of the predictions are above 0.9.}
  \label{relative}%
\end{table}%

In our prediction time period, though the observed real price range is large from $\$$ 775.98  to $\$$19498.68,
the predicted values based on our proposed prediction model well capture the trend of the real bitcoin price, as in figure  \ref{RA}(a).
As expected, the prediction performance of the proposed  model deteriorates as the observed real data series  is skyrocketing, as in November and December of 2017 in figure  \ref{RA}(a). However, through statistical analysis of the prediction results, of the 61 days in the last two months of 2017, on 25 days and 37 days relative accuracies were more than 0.8 and 0.7, respectively.

The overall average prediction accuracy of 362 consecutive days in 2017 can reach 0.82.  All the prediction results are shown in figure  \ref{RA}(b).  Of the 362 consecutive days in 2017, 82\%, 65\% and 37\% percent of the days achieve an accuracy of above 0.7, 0.8 and 0.9,  respectively. These statistics for the prediction results are summarized in table \ref{relative}.


\section{Discussions}
\label{end}
In this paper, a PDE model is developed
for bitcoin price prediction based on daily bitcoin transaction volumes and Google Trends Index. The average prediction accuracy (measured by relative accuracy) of our model is 0.82 for 362 consecutive days in 2017. Because of different datasets, it may be not comparable with other works. Nevertheless,  Kurbucz \cite{kurbucz2019predicting} achieves an accuracy of approximately 60.05\% during daily price movement classifications between November 25, 2016 and February 5, 2018. Jiang et al. \cite{jang2017empirical} obtains an acceptable prediction accuracy through selecting different relevant features of blockchain information and comparing the Bayesian neural network with benchmark models on modeling.


Our work differs from the previous prediction models with chainlets.
Akcora et al.  \cite{akcora2018forecasting} introduces ``chainlet" on Blockchain for bitcoin price prediction, but they focus only on the effects of certain types of chainlets on bitcoin price. Our proposed PDE model
emphasizes the combined effects from all the different types of chainlet clusters.  Further, the continuous model describes the influence of these chainlet clusters over time.

In addition, our PDE model differs from our previous  PDE models \cite{Haiyan2019book,wang2018prediction, wang2016regional} on social networks for predicting information diffusion, air pollution of 189 cities in China and influenza prevalence.  In this paper, our PDE model is developed  to capture the combined effect of chainlets from the Bitcoin transaction network and their influence on bitcoin price. In particular, the Google Trends Index is incorporated in the model to reflect the effect of market sentimental. Unlike the base linear or logistic models in \cite{Haiyan2019book,wang2018prediction, wang2016regional}, the PDE model in this paper has additional terms to describe the spatial heterogeneity of chainlet clusters. As a result, chainlet clusters with larger trading volumes do not necessarily have more influence on the bitcoin price. In fact, the prediction of  bitcoin price is based on the combined influence of all chainlet clusters.  
%



 \section*{Acknowledgements}
YW was partially supported by the Humanities and Social Sciences Research of the Ministry of
Education of China (18YJCZH184), the National Social Science Fund of China (19CGL002), the Natural Science Foundation of Tianjin (19JCQNJC14800),
China Postdoctoral Science Foundation (2018M640232).


\end{document}